\documentclass[amsmath,amssymb,prl,aps,twocolumn,mathbbm,superscriptaddress]{revtex4}
\usepackage{amssymb}
\usepackage{graphicx}
\usepackage{times}
\usepackage{subfigure}
\usepackage{color}
\usepackage{bbm}
\usepackage{mathtools}
\usepackage[usenames,dvipsnames]{xcolor}
\usepackage[colorlinks=true,citecolor=Cerulean,linkcolor=RubineRed,urlcolor=Cerulean]{hyperref}

\DeclarePairedDelimiter\floor{\lfloor}{\rfloor}

\begin{document}
\title{Digital Quantum Simulation of Minimal AdS/CFT}

\author{L. Garc{\'i}a-{\'A}lvarez}
\address{Department of Physical Chemistry, University of the Basque Country UPV/EHU, Apartado 644, E-48080 Bilbao, Spain}

\author{I. L. Egusquiza}
\affiliation{Department of Theoretical Physics and History of Science, University of the Basque Country UPV/EHU, Apartado 644, E-48080 Bilbao, Spain}

\author{L. Lamata}
\affiliation{Department of Physical Chemistry, University of the Basque Country UPV/EHU, Apartado 644, E-48080 Bilbao, Spain}

\author{A. del Campo}
\affiliation{Department of Physics, University of Massachusetts, Boston, MA 02125, USA}

\author{J. Sonner}
\affiliation{Department of Theoretical Physics, University of Geneva, 25 quai Ernest-Ansermet, 1214 Gen\`eve 4, Switzerland}

\author{E. Solano}
\address{Department of Physical Chemistry, University of the Basque Country UPV/EHU, Apartado 644, E-48080 Bilbao, Spain}
\address{IKERBASQUE, Basque Foundation for Science, Maria Diaz de Haro 3, 48013 Bilbao, Spain}

\date{\today}

\begin{abstract}
We propose the digital quantum simulation of a minimal AdS/CFT model in controllable quantum platforms. We consider the Sachdev-Ye-Kitaev model describing interacting Majorana fermions with randomly distributed all-to-all couplings, encoding nonlocal fermionic operators onto qubits to efficiently implement their dynamics via digital techniques. Moreover, we also give a method for probing non-equilibrium dynamics and the scrambling of information. Finally, our approach serves as a protocol for reproducing a simplified low-dimensional model of quantum gravity in advanced quantum platforms as trapped ions and superconducting circuits.
\end{abstract}

\maketitle

Holographic duality~\cite{Maldacena:1997re} posits the equivalence, subject to certain conditions, of quantum gravity and ordinary quantum field theories. The most celebrated such correspondence is conjectured to exist between ${\cal N}=4$ supersymmetric Yang-Mills theory in four dimensions and type IIB string theory on $AdS_5\times S^5$.  Such dualities offer the exciting prospect of probing quantum gravity effects by studying the well-defined equivalent quantum field theory. Nevertheless, this is still a hard problem because the semiclassical gravity regime is located at strong coupling and for a large number of local degrees of freedom $N\gg 1$. Furthermore, a fully non-perturbative understanding of the dual field theory is likely necessary in order to resolve the most puzzling aspects of quantum black holes, such as the famous information loss paradox~\cite{Hawking:1976ra}. We may therefore opt for studying the dual field theory on the lattice, by rewriting the problem in terms of a quantum many-body system suitable for simulation on a classical computer~\cite{Catterall:2007fp,Hanada:2013rga}. Even this powerful technique faces important challenges and limitations, such as the sign problem~\cite{Troyer:2004ge}, and the inapplicability of Euclidean lattice methods for intrinsically Lorentzian physics. It is precisely the latter kind of problem one needs to understand in order to describe black hole formation~\cite{Anous:2016kss} and evaporation. 

It is essential to develop alternative avenues of dealing with strongly coupled quantum many-body systems; both for their own sake, as well as with an eye on quantum gravity. As pointed out originally by Feynman~\cite{Feynman82}, quantum systems themselves are vastly more computationally efficient at solving many-body Hamiltonians than classical computer simulations. With the recent advent of quantum technologies~\cite{IonsReview,CircuitsReview1,CircuitsReview2,AtomsReview,PhotonsReview}, it is then natural to consider multiqubit systems that encode a dual gravity theory via quantum simulation. Currently, four-dimensional gauge theories  such as the aforementioned ${\cal N}=4$ theory appear out of reach (see, however, \cite{wiese2014towards} for work on QCD in this context). Instead, we start by looking elsewhere for simpler models which nevertheless have a holographic interpretation.

In this Letter, we propose the digital quantum simulation of the simplest known AdS/CFT duality, namely the Sachdev-Ye-Kitaev (SYK) model~\cite{AK15,Sachdev:2015efa,Danshita:2016aa}. We consider different variants of the model, two in terms of Majorana fermions, and two with complex fermions. We then propose  digital quantum algorithms for simulating the SYK quantum dynamics, and protocols to test non-equilibrium aspects such as scrambling. In particular, out of time order (OTO) four-point correlation functions, $\langle W^{\dagger}(t) V^{\dagger}(0) W(t)V(0)\rangle$, for initially commuting unitaries $W$ and $V$. Subsequently, we discuss the feasibility and implementation of our proposal in suitable quantum platforms such as trapped ions and superconducting circuits.

{{\it The holographic model.---}} The SYK model, in one of its simplest incarnations, is governed by the quenched-disorder Hamiltonian 
\begin{equation}\label{eq.SYKHamiltonian}
H= \frac{1}{4\cdot 4!}\sum_{i,j,k,l=1}^NJ_{ijkl}\chi_i \chi_j \chi_k\chi_l\,,
\end{equation}
where $\chi_i$ are Majorana fermions with $\{ \chi_i,\chi_j\}=2\delta_{ij}$, located on a lattice of $N$ sites and interacting via all-to-all couplings $J_{ijkl}$, sampled from a random distribution that is usually taken to be Gaussian with variance $\frac{3!J^2}{N^3}$. While similar models are common in the study of spin glasses~\cite{bray1980replica}, Hamiltonian~(\ref{eq.SYKHamiltonian}) lacks a spin-glass phase at low temperatures, making possible its holographic interpretation~\cite{polchinski2016spectrum}. Moreover, it has a number of striking features~\cite{AK15,Sachdev:2015efa,polchinski2016spectrum,maldacena2016comments}, beginning with its solvability in the limit of large $N$ and at strong coupling $\beta J\gg 1$, characterised by an approximate conformal symmetry. Furthermore, it exhibits maximally chaotic behaviour~\cite{AK15}, in the sense that the Lyapunov exponent $\lambda$, as extracted from a certain out-of-time order four-point function, saturates the bound $\lambda \ge 2\pi/\beta$~\cite{maldacena2015bound}. These features strongly suggest that the SYK model has a holographic interpretation in terms of an $NAdS_2$ (near-extremal $AdS_2$) theory of gravity~\cite{AK15,maldacena2016comments}.

Here we aim at the quantum simulation of quantum field theories with holographic duals, naturally starting with the simplest SYK model. Such simulations give us a way of solving the theory in any range of the coupling and for finite $N$, providing a realisation of a minimal quantum gravity model in the laboratory. From a theoretical perspective, analog gravity faces severe challenges~\cite{weinberg1980limits,marolf2015emergent}: any nonlinear gravity theory emerging from some local non-gravitational ``substrate'' will necessarily have its bulk dynamics entirely frozen. In other words, its bulk degrees of freedom may be entirely disregarded. The restrictions of ~\cite{weinberg1980limits,marolf2015emergent} are avoided in holographically emergent gravity, which is the path we pursue here.

{{\it SYK models.---}} The SYK model described by Eq.~(\ref{eq.SYKHamiltonian}) blends quantum gravity in a tractable fermionic Hamiltonian. One can also consider an alternative variant model in terms of complex spinless fermions capturing the same physics in the large $N$ limit, but in principle behaving differently for finite size $N$. To relate both models, we take $N=2n$, since two Majorana fermions provide us with one complex spinless fermion, and consider the Hamiltonian
\begin{equation}
H_{\rm{c}} = \frac{1}{(2 n)^{3/2}} \sum_{i,j,k,\ell=1}^n J_{ij;k\ell} \, c_i^\dagger c_j^\dagger c_k^{\vphantom \dagger} c_\ell^{\vphantom \dagger} 
- \mu \sum_{i} c_i^\dagger c_i^{\vphantom \dagger}, \label{H}
\end{equation}
with $\{c_i, c_j\}= 0$, $\{c_i,c_j^\dagger\} = \delta_{ij}$, and $\mu$ a chemical potential, while independent Gaussian random couplings $J_{ij;k\ell}$, are complex, with zero mean and such that
\begin{align}
J_{ji;k\ell} = - J_{ij;k\ell} \quad , \quad
J_{ij;\ell k} &= - J_{ij;k\ell}, \nonumber \\
J_{k\ell;ij} = J_{ij;k\ell}^\ast \quad , \quad
\overline{|J_{ij;k\ell}|^2} &= J^2 .
\end{align} 
Notice the different normalisation of  the coefficients: in the Majorana models the \( N^{-3/2} \) factor is in the variance of the coefficients, while here, in the complex fermion case, it has been taken out as a global factor.

We analyse the interaction terms appearing in both previous models for a subsequent treatment in a digital quantum simulation. In the model of Eq.~(\ref{eq.SYKHamiltonian}), with Majorana fermions, we identify two kinds of interaction terms: (i) $\chi_i \chi_j \chi_k \chi_l$ if all indices are distinct, and (ii) $\chi_i \chi_j$ if two subindices or three subindices coincide.  The case of all subindices being equal leads to a global phase in the evolution, which does not affect the dynamics simulation. Without loss of generality, we arrange terms such that $i>j>k>l$, where we have grouped instances with the same subindices by redefining the coupling constants  {(see Supplemental Material)}. Then, the fermionic interaction term count reads
\begin{align}
\text{(i)} \quad & \chi_i \chi_j \chi_k \chi_l  \text{ : }  \frac{2}{3}n^4 -2n^{3} + \frac{11}{6} n^{2} -\frac{1}{2} n \ , \nonumber \\ 
\text{(ii)} \quad & \chi_i \chi_j  \text{ : }  2n^2-n \ .\nonumber
\end{align}
Interaction terms of type (ii) cluster all Majorana fermionic terms with two or three matching subindices  {(see Supplemental Material)}.

Because of this separation, we also consider a different model with only type (i) terms, with the same large $N$ behaviour~\cite{maldacena2016comments}. Its implementation is straightforward given that of Eq.~(\ref{eq.SYKHamiltonian}), by restricting us to the simulation of the terms in case.

Secondly, the model formulated in terms of spinless complex fermions deals in principle with $(n^4 - 2n^3 +n^2)/4$ summands, which are classified in different kinds of interactions as (i) $c_i^\dagger c_j^\dagger c_k^{\vphantom \dagger} c_\ell^{\vphantom \dagger}$ with all indices distinct, (ii) $c_i^\dagger c_j^\dagger c_j^{\vphantom \dagger} c_\ell^{\vphantom \dagger}$ with $i$, $j$, and $\ell$ different from each other, and (iii) $c_i^\dagger c_j^\dagger c_j^{\vphantom \dagger} c_i^{\vphantom \dagger}$ with $i\neq j$. The number of terms to be simulated for each type of interaction is
\begin{align}
\text{(i)} \quad & c_i^\dagger c_j^\dagger c_k^{\vphantom \dagger} c_\ell^{\vphantom \dagger}  \text{ : } \frac{1}{4}n^4 -\frac{3}{2} n^3 +\frac{11}{4}n^2 -\frac{3}{2}n \ ,
\nonumber \\ 
\text{(ii)} \quad & c_i^\dagger c_j^\dagger c_j^{\vphantom \dagger} c_\ell^{\vphantom \dagger}  \text{ : } n^3 -3 n^2 + 2n \ ,
\nonumber \\ 
\text{(iii)} \quad & c_i^\dagger c_j^\dagger c_j^{\vphantom \dagger} c_i^{\vphantom \dagger}  \text{ : } \frac{1}{2}n^2 -\frac{1}{2}n \ ,
\nonumber \\
\text{(iv)} \quad & c_i^\dagger c_i^{\vphantom \dagger}  \text{ : } n \ . \nonumber
\end{align}
The indices in type (i) have been restricted to $i>j$ and $k>l$, thus reducing the number of terms to be simulated.

There is a straightforward variation of the model with the same holographic interpretation at large $N$: consider the couplings $J_{ij;k\ell}$ as purely real numbers. 

{\textit{Algorithm for quantum simulation of SYK models.---}}
The digital quantum simulation of the dynamics of the SYK models will involve fermionic operators, either Majorana or complex.  A quantum algorithm, consisting in a sequence of quantum gates on qubits, requires encoding fermionic operators into spin-$1/2$ operators. This is achieved via the Jordan--Wigner transformation~\cite{JordanWigner} from spinless complex fermion operators to spin-$1/2$ operators, $c_i^\dagger = ( \prod_{j=1}^{i-1} \sigma^z_j ) \sigma^{+}_i$. One can define a set of $2n$ Majorana fermions as $\chi_{2j-1} = e^{i\phi} c_j + e^{-i\phi} c_j^{\dagger}$, $\chi_{2j} = -i(e^{i\phi} c_j - e^{-i\phi} c_j^{\dagger})$, and codify the Majorana fermionic operators as $\chi_{2n-1} = ( \prod_{j=1}^{n-1} \sigma^z_j) \sigma^x_n$ and $\chi_{2n} = ( \prod_{j=1}^{n-1} \sigma^z_j )\sigma^y_n$, with $\{\chi_i,\chi_j\}=2\delta_{ij}$.

Interactions in the Majorana models lead to two kinds of fermionic terms, that we rewrite in terms of spin degrees of freedom as 
\begin{equation} \label{generalint}
\chi_i \chi_j \chi_k \chi_l = \sigma^{\alpha_i}_{\tilde{i}} \left(\prod_{m = \tilde{j}}^{\tilde{i}-1} \sigma^z_m \right) \sigma^{\alpha_j}_{\tilde{j}} \sigma^{\alpha_k}_{\tilde{k}} \left(\prod_{m = \tilde{l}}^{\tilde{k}-1} \sigma^z_m \right) \sigma^{\alpha_l}_{\tilde{l}},
\end{equation}
and
\begin{equation}
\chi_i \chi_j = \sigma^{\alpha_i}_{\tilde{i}} \left(\prod_{m = \tilde{j}}^{\tilde{i}-1} \sigma^z_m \right) \sigma^{\alpha_j}_{\tilde{j}},
\end{equation}
where $i>j>k>l$. Here, we define the tilded variables with the floor function as
\begin{equation}
\tilde{x} = \floor*{\frac{x+1}{2}}= \max \Big\{m \in \mathbb{Z} \ | \ m \leq \frac{x+1}{2} \Big \},
\end{equation}
and the $\alpha_n$ labels correspond to
\begin{equation}
\alpha_n = \left \{ \begin{matrix} x & \mbox{if }n\mbox{ is odd}
\\ y & \mbox{if }n\mbox{ is even}\end{matrix}\right. \quad .
\end{equation}
Among the resulting spin interaction terms, the most general and complex form corresponds to that shown in Eq.~(\ref{generalint}). In some specific cases of combination of indices the expression is simplified  {(see Supplemental Material)}.

Let us now consider the model with complex spinless fermions. The interaction terms can be mapped as above to spin interactions via the Jordan--Wigner transformation. Thus, the interaction terms of type (i) of this model are expressed as
\begin{equation}
c_i^\dagger c_j^\dagger c_k^{\vphantom \dagger} c_\ell^{\vphantom \dagger} = \beta \left(\prod_{\xi= \zeta_1 +1}^{\zeta_2 -1} \sigma^z_\xi \right) \left(\prod_{\xi= \zeta_3 +1}^{\zeta_4 -1} \sigma^z_\xi \right) \sigma^+_i \sigma^+_j \sigma^-_k \sigma^-_\ell \ ,
\end{equation}
where $\{\zeta_1,\zeta_2,\zeta_3,\zeta_4\} = \{i,j,k,\ell\}$ as sets,  $\zeta_1< \zeta_2< \zeta_3<\zeta_4$, and $\beta= \text{sign}(i-j) \text{sign}(\ell-k)$. For the sake of simplicity in the quantum simulation, we have only taken into account the terms such that $i>j$ and $k>l$, wherefore  $\beta = -1$.

The interaction terms of type (ii), (iii) and (iv) can also be mapped to spin interactions as
\begin{align}
\text{(ii)} & \quad  c_i^\dagger n_j c_k^{\vphantom \dagger} =  -\frac{1}{2}\left(\prod_{\xi=\zeta_1+1}^{\zeta_2-1} \sigma^z_\xi \right) (\sigma^{z}_j +1) \sigma^{+}_i \sigma^{-}_k , \\
\text{(iii)} & \quad n_i n_j =  \frac{1}{4} \left( 1 + \sigma^{z}_i + \sigma^{z}_j + \sigma^{z}_i \sigma^{z}_j \right) , \\ 
\text{(iv)} & \quad n_i =  \frac{1}{2} (1 + \sigma^{z}_i) \ ,
\end{align}
where $\{\zeta_1,\zeta_2\} = \{i,k\}$, again as sets, and $\zeta_1< \zeta_2$. It is still possible to reduce the number of interaction terms by considering the properties of coefficients $J_{ij;k\ell}$  {(see Supplemental Material)}.

These spin Hamiltonians consist of a sum of spin interactions $H = \sum_{i}^m H_i$, with $H_i$ a many-body spin interaction. Obtaining the exact evolution is a difficult problem to deal with a purely analog quantum simulation in any quantum platform. Nevertheless, it is possible to handle individually each spin interaction in digital quantum simulations~\cite{Lloyd1073} by decomposing the evolution operator in a Trotter--Suzuki product formula,
\begin{equation}
e^{-i H t} = \left(\prod_{j=1}^m e^{-i H_j t/s}\right)^s + \sum_{i<j} \frac{[H_i,H_j] t^2}{2s} + O(J^3t^3/s^2) .
\end{equation}
This expression approximates the dynamics for time $t$ to an accuracy $\epsilon$ of the order of $J^2 t^2/s$. We note that for each non-zero commutator, $[H_i,H_j]\neq 0$, there is a decrease in  accuracy. In the worst case scenario, where all the commutators differ from zero, there will be a reduction of accuracy given by the factor $\binom {m} {2}$, with $m$ the number of interaction terms $H_i$ in the Hamiltonian.

The complexity of the algorithm, i.e. the number of gates required for the dynamics simulation, grows polynomially with the number of fermions $N$. The coarsest evaluation suggests that achieving an accuracy \( \epsilon \) over an evolution time \( t \) will require a number of gates \( m\times \binom{m}{2}\times J^2 t^2/\epsilon \). In the case at hand, there will be \( m \sim O(N^4) \) spin interactions. If each interaction is given by \( O(1) \) gates, the number of gates for accuracy \(\epsilon\) over a time \( t \) will be \( O(N^{12}) \).
In fact, the number of non-zero commutators is of order \( O(N^6) \), rather than \( m^2\sim O(N^8) \), thus bringing the number of gates down to \( O(N^{10}) \).
As usual, higher order Trotter--Suzuki decompositions will improve the accuracy of our simulation.

{\textit{Protocol for time inversion.---}}
In order to probe the non-equilibrium behaviour of the SYK, and more specifically, the  dynamics of scrambling~\cite{maldacena2015bound} in terms of OTO functions, it will be necessary to reverse the evolution via a time inversion operation. This can be achieved by constructing an operator $U(-t)$, where $U(t)$ denotes the time-evolution operator of the system. Since the models are described by time independent Hamiltonians, we need only to reverse the sign of all the couplings. Alternatively, we show that time inversion can also be implemented without explicitly engineering the algorithm for $U(-t)$. We consider an additional ancilla qubit $Q_C$, which controls the direction of the SYK time evolution allowing us to implement both $U(t)$ and $U(-t)$. The evolution for the pair controlling ancilla and system is governed by the Hamiltonian $H_{CS}= \sigma_{C}^z H_S$, where $\sigma_{C}^z$ acts on the ancilla qubit, and $H_S$ is the system Hamiltonian. The complete Hamiltonian consists in a tensor that acts on the ancilla-system product state as
\begin{align}
H_{CS}(\alpha |e\rangle + \beta |g\rangle) |\psi \rangle &= \sigma_{C}^z(\alpha |e\rangle + \beta |g\rangle) H_S |\psi \rangle \nonumber \\
&= \alpha |e\rangle H_S |\psi \rangle - \beta |g\rangle H_S |\psi \rangle ,
\end{align}
with $\alpha |e\rangle + \beta |g\rangle$ and $|\psi \rangle$ the ancilla and system states, respectively. In general, one may consider $n$ applications of the Hamiltonian 
\begin{align}
H^n_{CS}(\alpha |e\rangle &+ \beta |g\rangle) |\psi \rangle = (\sigma_{C}^z)^n(\alpha |e\rangle + \beta |g\rangle) H^n_S |\psi \rangle \nonumber \\
&= \alpha |e\rangle H_S |\psi \rangle + \beta |g\rangle (-1)^n H_S |\psi \rangle ,
\end{align}
which leads to a time evolution for an initial product state $|\Psi \rangle = (\alpha |e\rangle + \beta |g\rangle) |\psi \rangle$ described by
\begin{align}
&U_{CS}(t) |\Psi \rangle = \exp\left(- i H_{CS} t\right)|\Psi \rangle = \sum_n \frac{(- i H_{CS} t)^n}{n!} |\Psi \rangle \nonumber \\ 
&= \alpha |e\rangle \sum_n \frac{(- i H_{S} t)^n}{n!} |\psi \rangle + \beta |g\rangle \sum_n \frac{(-1)^n(-i H_{S} t)^n}{n!} |\psi \rangle\nonumber \\
&= \alpha |e\rangle U(t) |\psi \rangle + \beta |g\rangle U(-t) |\psi \rangle .
\end{align}
Here, the controlling ancilla qubit decides the direction of the evolution. The scrambling four-point function for operators \( V_S \)  and \( W_S \)  is thus computable by the sequence
\begin{equation}
\sigma_C^x U_{CS}(t) \sigma_C^x W^{\dagger}_S U_{CS}(t) V^{\dagger}_S \sigma_C^x U_{CS}(t) \sigma_C^x W_S U_{CS}(t) V_S 
\end{equation}
and its expectation value over the state $|e\rangle\otimes|\psi\rangle$. We notice that operators labeled by $S$, all of them feasible in our proposed scheme, are only applied to the system $S$ (For an analogous construction, see~\cite{Grover16}).

{\textit{Protocol for correlation measurements.---}}
We consider an alternative and efficient protocol for determining $n$-time correlation functions~\cite{Julen14}. In this alternative method, an ancillary qubit $Q_A$ encodes a correlation function by means of controlled operations. This approach is particularly effective for analog quantum simulation of the evolution, but it is also applicable to digitally synthesised quantum evolutions.

Let us consider a set of unitaries \(\{V_{i}\}_{i=0}^n \) acting on the system. We aim at computing the correlation function \( \langle V_n(t_n)V_{n-1}(t_{n-1})\cdots V_1(t_1)V_0(0)\rangle \). Let us assume that we know how to implement in the laboratory the unitaries \( \tilde{V}_i = |0\rangle\langle0|_A I +|1\rangle\langle1|_A V_{i} \), which act on the ancilla and the system . The ancilla $Q_A$ and the system evolve with \( I_A U(t) \), where \( U(t) \) is the unitary evolution operator for the system $S$. We consider now the following protocol:
a) Prepare the system in the state of interest, \( |\Psi\rangle \), and the ancilla in the state \( \left(|0\rangle+|1\rangle\right)/ \sqrt{2} \). Set counter \( k \) to 0. Define $t_0 = 0$.
b) Apply the controlled unitary \(\tilde{V}_k \).
c) Evolve a time \( t_{k+1} - t_k \).
d) If \( k<n \), then  advance \( k \) by 1 and go to step b). Else, measure \( \sigma_x \) and \( \sigma_y \) of the ancilla $Q_A$, which completes the protocol. This leads to the desired measurement of the $n$-time correlation function
\[ \frac{1}{2}\left( \langle \sigma_x\rangle + i \langle \sigma_y\rangle\right)=  \frac{1}{2} \langle V_n(t_n)V_{n-1}(t_{n-1})\cdots V_1(t_1)V_0(0)\rangle . \]

For the particular case of the scrambling four-point correlation function $\langle W^{\dagger}_S(t) V^{\dagger}_S(0) W_S(t)V_S(0)\rangle$, the set of unitaries to be considered in the protocol is \(\{W^{\dagger}_S,V^{\dagger}_S, W_S,V_S \}\), with corresponding times $t_1=t$, $t_2=0$, and $t_3=t$. A similar approach in this context has been recently proposed~\cite{Swingle:2016aa}. Note that in order to evolve the system one requires time inversion, since $t_2 < t_1$, for which case the protocol introduced previously can be used.

 \begin{figure*}[t]
\centering
(a)
\includegraphics[angle=0, height=0.3\textwidth]{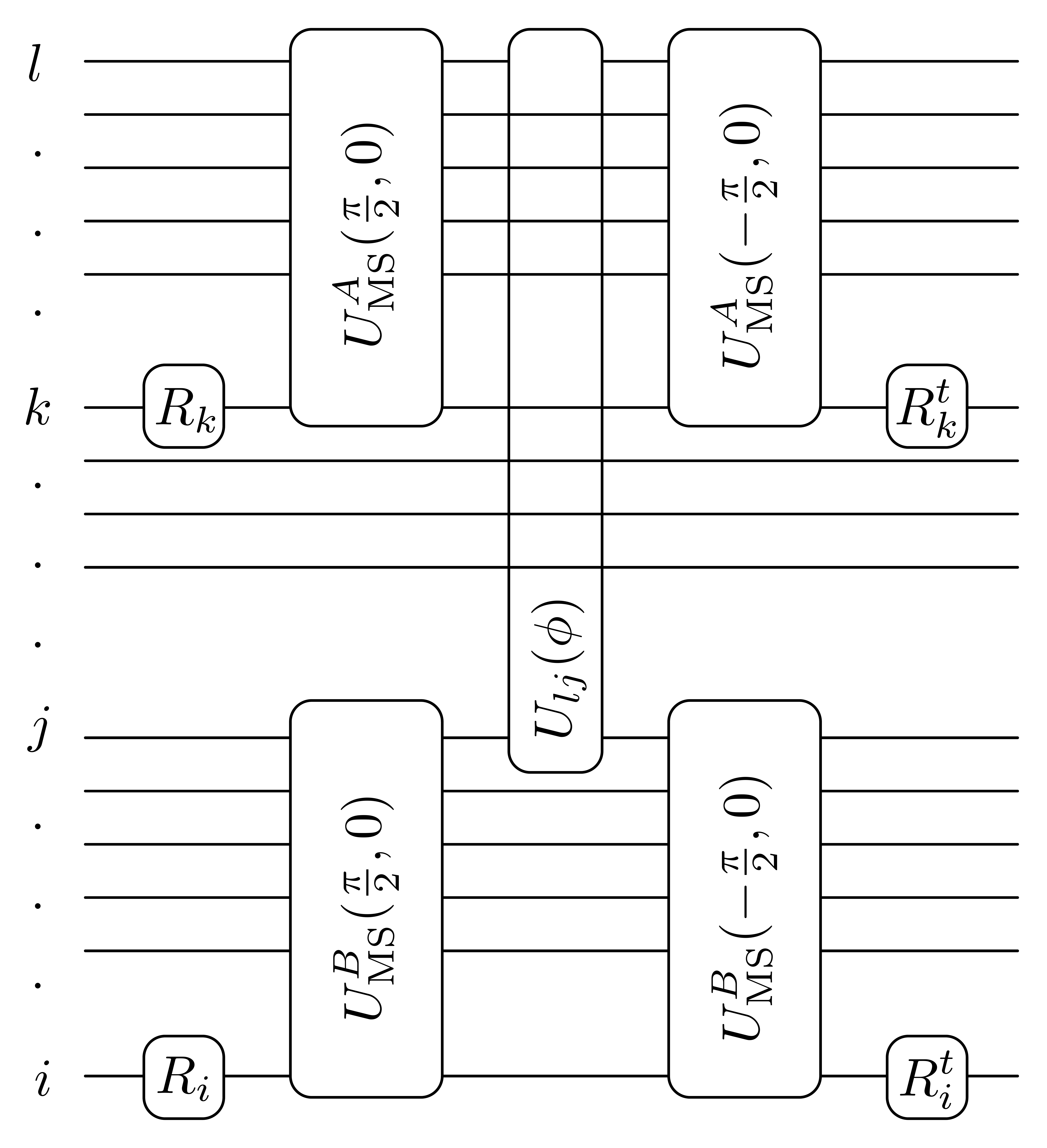}\hspace{1cm}
(b)
\includegraphics[angle=0, height=0.3\textwidth]{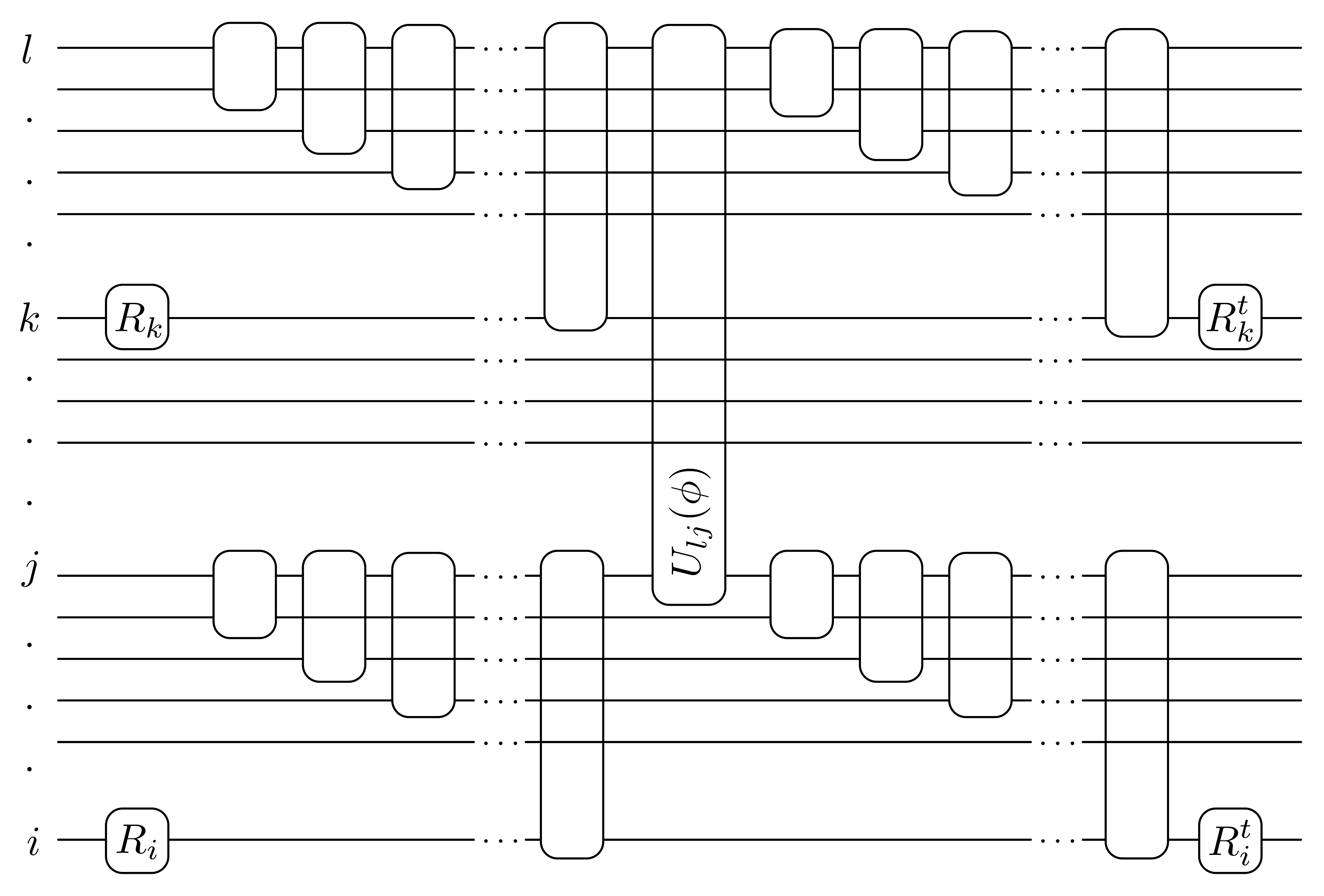}
	\caption{{\bf Engineering many-body interactions.}   (a) Trapped-ion qubits: Operation sequence of single-qubit and multiqubit gates, inside a Trotter step, acting on trapped-ion qubits to generate a generic interaction term (\ref{Hiterm}). The single-qubit rotations $R_i$ and $R_k$ act on qubits $i$ and $k$, respectively, and the phase $\phi$ of the two-qubit entangling gate, $U_{ij}(\phi)$, must be chosen adequately in order to produce the desired combination of $\alpha_i \alpha_j \alpha_k \alpha_l$ in the interaction. (b) Superconducting circuits: We consider sets of two-qubit gates and their inverses, which involve qubits $l$ and $j$ with the rest of the qubits included in the $\sigma^z$ strings of the interaction. Thus, a set of $n$ two-qubit gates takes on the role of the M\o lmer-S\o rensen gate in the trapped-ion protocol. Note that two-qubit gates between distant qubits may be performed by a set of SWAP gates and an entangling gate between nearest-neighbour qubits.
		\label{fig1}}
\end{figure*}

{\textit{Protocol  for state initialisation.---}}
Scrambling depends on the Hamiltonian structure for typical initial states. It is possible to prepare thermal states on a quantum computer following existing methods in the literature~\cite{EisertTS12,WilhelmTS16}. Moreover, it is also possible to analyse scrambling for explicitly known initial states. In order to do that, one uses again the Jordan--Wigner transformation to encode fermionic states in qubit states. Let us now give some  examples.

In the case of the spinless complex fermions, the mapping $c_i^\dagger = (\prod_{j=1}^{i-1} \sigma^z_j) \sigma^{+}_i$ indicates that the fermionic state with zero excitations, $|\vec{0}\rangle$, corresponds to the product state of all the qubits in their ground state, $|g_1, \dots , g_n \rangle$, since
\begin{equation}
c_i |\vec{0}\rangle = \sigma^{-}_i \left(\prod_{j=1}^{i-1} \sigma^{z}_j\right) |\vec{0}\rangle = 0 \quad \forall i \Leftrightarrow |\vec{0}\rangle = |g_1, \dots , g_n \rangle .
\end{equation}

Once this identification is known, one may consider the action of a set of creation fermionic operators on $|\vec{0} \rangle$ by acting with their spin correspondents on $|g_1, \dots , g_n \rangle$. This way, a state with a certain number of excitations in localised fermionic sites can be constructed. In particular, the zero fermion excitation state above, which is a maximal total spin one.

{\textit{Implementation in quantum platforms.---}}
In order to simulate the interactions, we consider a generic term
\begin{equation}
\label{Hiterm}
H_i = \left(\prod_{m = l}^{k-1} \sigma^z_m \right) \left(\prod_{m = j}^{i-1} \sigma^z_m \right) \sigma^{\alpha_i}_{i} \sigma^{\alpha_j}_{j} \sigma^{\alpha_k}_{k} \sigma^{\alpha_l}_{l},
\end{equation}
with any combination of $\alpha_i \alpha_j \alpha_k \alpha_l$, and two separated $\sigma^z$ strings. The simulation of the remaining interactions may be inferred from this technique, since this is the most general kind of spin interactions appearing in the Hamiltonian.

Each of these spin interactions will appear with a different coupling strength \( j_i J \), determined by a realisation of the random couplings. Notice that the protocol will in principle run for several instantiations of the couplings. This, however, might  be not necessary for large $N$, due to self-averaging.

The time evolution generated by this term is described by $U(t) = \exp(-i j_i J H_i t)$. Different protocols for the quantum simulation of this unitary operator are platform dependent. In particular, the coupling strength \( j_i J  \) will be realised by control of phases in the gates in both the trapped ion and the superconducting circuits schemes. 

{\textit{i. Trapped Ions.}}
The platform of trapped ions has been a workhorse of quantum simulations for some time now~\cite{lanyon2011universal,Martinez:2016aa}.
The efficient implementation of exponentials of tensor products of $j$ Pauli matrices in trapped ions relies on the implementation of the M{\o}lmer-S{\o}rensen gate, involving $M$ qubits and local rotations~\cite{1367-2630-13-8-085007}. We consider the standard expression $U_{\rm MS} (\theta,\varphi)=\exp[-i\theta(\cos(\varphi) S^x +\sin(\varphi) S^y)^2/4]$, and $S^{x,y}=\sum_{j=0}^M \sigma^{x,y}_j$. In order to make contact with the literature in trapped ions, we use in this paragraph a different basis, such that $|g\rangle_z, |e\rangle_z \rightarrow |\tilde{g}\rangle_x, |\tilde{e}\rangle_x$, and the corresponding mapping \( \sigma^z\rightarrow \tilde{\sigma}^x\), \( \sigma^x\rightarrow \tilde{\sigma}^y\), \( \sigma^y \rightarrow \tilde{\sigma}^z \).  The tilded objects refer to the ion implementation, while the untilded ones come from our algorithms. Having made this distinction, in fact above and in what follows we obviate the tildes.

Since our quantum algorithm requires the simulation of terms with two disjoint Jordan--Wigner  tails, we  consider gates $U^{A}_{\rm MS}$ and $U^{B}_{\rm MS}$ associated with the disjoint collective operators $S_A^{x,y}=\sum_{\ell=l}^{l+M} \sigma^{x,y}_{\ell}$ and $S_B^{x,y}=\sum_{\ell=j}^{j+K} \sigma^{x,y}_{\ell}$, respectively, and the entangling gate $U_{lj} (\phi) = \exp(i\phi \sigma_l^{\alpha} \sigma_j^{\beta})$. We propose then the following step shown in Fig. \ref{fig1}a, that scales with $O(1)$ in the number of fermions $N$,
\begin{align} \label{step}
U =&\, U^{A}_{\rm MS} (-\tfrac{\pi}{2},0) U^{B}_{\rm MS} (-\tfrac{\pi}{2},0) U_{lj} (\phi) U^{B}_{\rm MS} (\tfrac{\pi}{2},0) U^{A}_{\rm MS} (\tfrac{\pi}{2},0) \nonumber \\ 
=& \exp\Big[i\phi \Big( \cos\left(\tfrac{\pi}{2} s_A^x\right) \cos\left(\tfrac{\pi}{2} s_B^x\right) \sigma_l^{\alpha} \sigma_j^{\beta} \nonumber \\
&+ \cos\left(\tfrac{\pi}{2} s_A^x\right) \sin\left(\tfrac{\pi}{2} s_B^x\right) \epsilon_{x\delta\beta} \sigma_l^{\alpha} \sigma_j^{\delta} \nonumber \\
&+ \sin\left(\tfrac{\pi}{2} s_A^x\right) \cos\left(\tfrac{\pi}{2} s_B^x\right) \epsilon_{x\gamma\alpha} \sigma_l^{\gamma} \sigma_j^{\beta} \nonumber \\
&+ \sin\left(\tfrac{\pi}{2} s_A^x\right) \sin\left(\tfrac{\pi}{2} s_B^x\right) \epsilon_{x\beta\delta} \epsilon_{x\alpha\gamma} \sigma_l^{\gamma} \sigma_j^{\delta} \Big)\Big] ,
\end{align}
with $s_A^{x}=\sum_{i=l+1}^{l+M} \sigma^{x}_i$, $s_B^{x}=\sum_{i=j+1}^{j+K} \sigma^{x}_i$, and \(\epsilon_{ \alpha \beta \gamma}\) the Levi--Civita symbol. For a generic operator $s^x = \sum_{i=1}^{n} \sigma^{x}_i$ acting on $n$ qubits, we take into account the identities
\begin{equation}
\cos\left(\tfrac{\pi}{2} s^x\right)= \left\{ \begin{array}{lcl}
             - \prod_{i=1}^{n} \sigma^{x}_i &  \text{for}  & n=4k-2, \ k \in \mathbb{N} \ ,\\
             \\ \prod_{i=1}^{n} \sigma^{x}_i & \text{for} & n=4k, \ k \in \mathbb{N} \ ,\\
             \\ 0 & \text{for} & n \ \text{odd}\ ,
             \end{array}
   \right.
\end{equation}
and
\begin{equation}
\sin\left(\tfrac{\pi}{2} s^x\right)= \left\{ \begin{array}{lcl}
             \prod_{i=1}^{n} \sigma^{x}_i &  \text{for}  & n=4k-3, \ k \in \mathbb{N} \ ,\\
             \\ - \prod_{i=1}^{n} \sigma^{x}_i & \text{for} & n=4k-1, \ k \in \mathbb{N} \ , \\
             \\ 0 & \text{for} & n \ \text{even}\ .
             \end{array}
   \right. 
\end{equation}
We observe now that we can engineer the desired terms where the choice of the intermediate entangling gate $U_{lj} (\phi)$ allows us to modify the resulting interaction term
\bigskip
\begin{align}
U = \left\{ \begin{array}{l}
             \exp \Big[ i\phi \big(a(M)\prod_{i=l+1}^{l+M} \sigma^{x}_i \big) \big(a(K)\prod_{k=j+1}^{j+K} \sigma^{x}_k \big) \sigma^{\alpha}_l \sigma^{\beta}_j \Big], \\
             \\ \exp \Big[ i\phi \big(b(M)\prod_{i=l+1}^{l+M} \sigma^{x}_i \big) \big(b(K)\prod_{k=j+1}^{j+K} \sigma^{x}_k \big) \\\times\epsilon_{x\beta\delta} \epsilon_{x\alpha\gamma} \sigma_l^{\gamma} \sigma_j^{\delta} \Big], \\
             \\ \exp \Big[ i\phi \big(a(M)\prod_{i=l+1}^{l+M} \sigma^{x}_i \big) \big(b(K)\prod_{k=j+1}^{j+K} \sigma^{x}_k \big) \\\times \epsilon_{x\delta\beta} \sigma_l^{\alpha} \sigma_j^{\delta} \Big], \\
             \\ \exp \Big[ i\phi \big(b(M)\prod_{i=l+1}^{l+M} \sigma^{x}_i \big) \big(a(K)\prod_{k=j+1}^{j+K} \sigma^{x}_k \big) \\\times \epsilon_{x\gamma\alpha} \sigma_l^{\gamma} \sigma_j^{\beta} \Big],
             \end{array}
   \right.
\end{align}
for the cases $M$ and $K$ even, $M$ and $K$ odd, $M$ even and $K$ odd, and $M$ odd and $K$ even, respectively, and with
\begin{equation}
a(n)= \left\{ \begin{array}{lcl}
             -1 &  \text{for}  & n=4k-2, \ k \in \mathbb{N}\ , \\
             \\ 1 & \text{for} & n=4k, \ k \in \mathbb{N}\ ;
             \end{array}
   \right.\nonumber
\end{equation}
\begin{equation}
b(n)= \left\{ \begin{array}{lcl}
             1 &  \text{for}  & n=4k-3, \ k \in \mathbb{N}\ , \\
             \\ -1 & \text{for} & n=4k-1, \ k \in \mathbb{N} \ .            
             \end{array}
   \right. 
\end{equation}

We note that for the generic and most complex interaction, Eq. (\ref{Hiterm}), the Jordan--Wigner $\sigma^z$ tails begin in sites $l$ and $j$, and end in sites $k$ and $i$, all corresponding to the ones of the four-body interaction. Up to now, we have achieved a many-body interaction involving $l$ and $j$ sites and two corresponding tails starting in those sites and ending in $l+M$ and $j+K$. The desired interaction can be easily achieved by considering that $l+M$ and $j+K$ correspond to $k$ and $i$, respectively, and by applying the corresponding rotations in the $k$-th and $i$-th qubits to obtained the desired Pauli matrices, as depicted in Fig. \ref{fig1}a.

{\textit{ii. Superconducting circuits.}}
The general framework of superconducting quantum processors is an extremely active area of research~\cite{houck2012chip,Barends:2015aa,PhysRevX.5.021027,Barends:2016aa,IBM}.
The translation of the previous protocol for a generic interaction to superconducting circuits can be immediately done by considering the application of the multiqubit M{\o}lmer-S{\o}rensen gate via a superconducting resonator~\cite{MezzacapoMS}. Alternatively, the decomposition of the basic step in Eq.~(\ref{step}) into single-qubit and two-qubit gates by breaking the M{\o}lmer-S{\o}rensen gate into $\binom{n}{2}$ two-qubit gates can be used in principle. We consider for the sake of simplicity, and without loss of generality, a single M{\o}lmer-S{\o}rensen gate associated with a collective operator $S^x$ of $n$ qubits, and the sequence of $U_{\rm MS} (\tfrac{\pi}{2},0) U_i(\phi) U_{\rm MS} (-\tfrac{\pi}{2},0)$, with $U_i(\phi)=\exp(i\phi\sigma_i^{\alpha})$  an intermediate single-gate. In this protocol, shown in Fig. \ref{fig1}b, if we decompose  $U_{\rm MS} (\tfrac{\pi}{2},0)$ and its inverse into two-qubit gates, we realise that the only operations that do not cancel out are those involving the $i$-th qubit, in which $U_i(\phi)$ is applied. This fact implies that instead of $\binom{n}{2}$ two-qubit gates per collective gate, we reduce the number of entangling gates to $n$ in the simulation of each Hamiltonian $H_i$, i.e. we have a scaling of $O(N)$ gates per interaction. 

We may consider not only linear arrays of qubits, but also bidimensional lattices~\cite{0295-5075-85-5-50007,Zohar:2016aa} in which the qubit connectivity increases with a qubit having four nearest neighbours. Thus, one can implement the Jordan--Wigner transformation as above while reducing the number of SWAP gates needed in the protocol. Another extension, which may be needed for the case of quantum field theories, is to consider digital-analog quantum simulations~\cite{CiracDQS,Laura14,Arrazola16}. In this manner, we can exploit the concept of complexity-simulating-complexity while merging digital and analog techniques in a complementary fashion.

{\textit{Conclusions.---}} We have proposed the digital quantum simulation of  Sachdev-Ye-Kitaev models as a minimal realisation of the AdS/CFT duality in quantum technologies. We encode the SYK nonlocal fermionic model onto a multiqubit system, and show how to efficiently simulate its dynamics with digital techniques and polynomial resources. We also provide a protocol for studying the non-equilibrium behaviour, including the scrambling of information. Our proposal could be implemented with state-of-the-art trapped ions and superconducting circuits, paving the way towards the realisation of minimal quantum gravity models in the laboratory, enhancing the toolbox of quantum simulations.

We acknowledge support from a UPV/EHU PhD grant, Spanish MINECO/FEDER FIS2015-69983-P, UPV/EHU UFI 11/55 and Project EHUA14/04,  Ram\'on y Cajal Grant RYC-2012-11391, and the John Templeton foundation. JS is supported by the Fonds National Suisse de la Recherche Scientifique (FNS) under grant number 200021 162796 and by the NCCR 51NF40-141869 ``The Mathematics of Physics" (SwissMAP).

\bibliography{biblio}

\begin{widetext}
\section*{SUPPLEMENTAL MATERIAL}

\section{SYK models}
We have optimised the number of terms to simulate in all models, i.e. with Majorana and complex fermions, by grouping interaction terms via the anticommutation relations of fermionic operators. These simplifications lead to new coupling coefficients resulting from linear combinations of the original coefficients of fermionic models, $J_{ij;k\ell}$ and $J_{ijkl}$. These new groupings of coefficients will need to be generated for the simulation, and implemented as phases in the quantum gates. Thus, we analyse which are independent random variables and their distribution in relation to those of the initial models.

\subsection{Details on Majorana models}
For the Majorana models we start with the case \( i>j>k>l \) (type (i) interaction terms), and group the interaction coupling constants as follows
\begin{equation}
\tilde{J}_{\alpha_4\alpha_3\alpha_2\alpha_1} = \sum_{\sigma(\alpha_4\alpha_3\alpha_2\alpha_1)}\text{sgn}(\sigma) J_{\sigma(\alpha_4\alpha_3\alpha_2\alpha_1)} = 4! J_{\alpha_4\alpha_3\alpha_2\alpha_1} \ ,
\end{equation}
where we have taken into account that, in this case, $J_{jklm} = \text{sgn}(\sigma) J_{\sigma(jklm)}$. A linear transformation of an stochastic variable, \( \eta= a \xi+ b \), induces a transformation of the probability density \( p_\eta(y)=(1/|a|) p_\xi[( y-b)/a] \), and this keeps us in the Gaussian family: if \( \xi\sim N( \mu, \sigma) \), then \( \eta\sim N(a\mu+b,|a| \sigma) \).  In our case, since \( J_{jklm}\sim N(0, \sqrt{6} J N^{-3/2} )\), we have  \( \tilde{J}_{jklm}\sim N(0, 4!\sqrt{6} J N^{-3/2}) \). Defining \( \tilde{J}=4! J \), we note that the new coefficients $\tilde{J}_{jklm}$ present the same structure as the initial ones, with a rescaled variance. 

The purpose of this simplification is to identify independent stochastic variables; clearly the constraints \( J_ {\alpha_4\alpha_3\alpha_2\alpha_1} = \mathrm{sign}\left( \sigma\right)J_{\sigma(\alpha_4\alpha_3\alpha_2\alpha_1)} \) signify that these are not independent. The \( \tilde{J} \) variables are symmetrised versions, and are independent.

Our first Majorana model under consideration includes only the four-body interactions above (type (i)), i.e. those with no repeated indices, that can be expressed in terms of \( \chi_i \chi_j \chi_k \chi_l \) with \( i>j>k>l \).

Let us now focus on terms with repeated indices, which give rise to type (ii) interaction terms. Taking them into account  as well as the previous ones provides us with a second Majorana model. Notice however that  the  physical properties in the limit of large \( N \) would not change. Whenever present, the coefficients \( J_{ijkl} \) with repeated indices would no longer possess the symmetry described above. For instance, coefficients of \( \chi_i \chi_j \) terms  (to which the four fermion operators with repeated indices reduce) should be purely imaginary: let the two fermion terms be written as the operator \( \mathcal{M}_2=\chi^T A \chi \), with \( \chi \) a column vector with entries the Majorana operators. The non-trivial elements correspond to the antisymmetric part of the matrix \( A \), so this is taken as antisymmetric. In order for this operator \( \mathcal{M}_2 \) to be hermitian one additional constraint is required: \( A \) must be hermitian. It follows that \( A \) must be purely imaginary. If we consider a model in which these terms are present, the coefficients \( A_{ij} \) with \( i>j \) will be taken from a Gaussian ensemble, \( A_{ij}\sim N(0, \sigma_A ) \). These coefficients will be independent of the previously introduced $\tilde{J}_{jklm}$. We give the variance as \( \sigma_A =  J_{A} /(2\sqrt{N}) \). Notice however that if we were to start from the complex fermion model, with or without chemical potential $\mu$, and rewrite it in terms of Majorana fermions, the coefficients of the two Majorana fermionic terms will not be independent of the four-point ones in that translated model. 

The previous presentation can be summarised as follows: we identify independent interaction terms. Those correspond to $i>j>k>l$, four-point interactions, and $i>j$, two point interactions. The coupling constants corresponding to these independent interaction terms have been presented above.  The number of independent terms of each kind, four-point or two point, is given by \( \binom{N}{4} \)  and \( \binom{N}{2} \), respectively.

%
%
%
\subsection{Details on spinless complex fermion models}
We now identify the number of independent interaction terms for the spinless complex fermion models, and the corresponding coefficients and their probability distribution. They are firstly rearranged as follows
\begin{equation}
\tilde{J}_{\alpha_2\alpha_1;\beta_2\beta_1} = \sum_{\sigma_1(\alpha_2\alpha_1);\sigma_2(\beta_2\beta_1)}\text{sgn}(\sigma_1) \text{sgn}(\sigma_2) J_{\sigma_1(\alpha_2\alpha_1);\sigma_2(\beta_2\beta_1)} = 4 J_{\alpha_2\alpha_1;\beta_2\beta_1} \ ,
\end{equation}
with  $\alpha_2>\alpha_1$ and $\beta_2>\beta_1$. Otherwise the tilded coefficients are defined to be zero.
This takes into account the fermionic symmetries of the creation and annihilation operators separately, and allows us to concentrate on the case \( \alpha_2> \alpha_1 \) and  \( \beta_2>  \beta_1 \), which will be useful for the quantum simulation of interaction terms of type (i), as we will see. 

Altogether there are, for \( n \) spinless fermions, \( \binom{n}{2}^2 \) terms of the form \( c^{\dag}_i c^{\dag}_j c_k c_l \) that satisfy the constraints \( i>j \) and \( k>l \). 

Now, let us connect with the four types presented in the main text. We identify first terms of type (i), for which there is no coincidence of indices, i.e. such that the additional constraints \( i\neq k \), \( i\neq l \), \( j\neq k \) and \( j\neq l \) hold, as separate from those cases in which there is coincidence.

Interaction terms of type (ii) have a coincidence of two indices with the other two distinct from each other and from the repeated one, that is, a generic interaction $c_i^\dagger n_j c_k^{\vphantom \dagger}$, with no ordering of indices imposed on them. The counting of combinations of three different indices taken from a set of $n$ is $\frac{n!}{(n-3)!} =  n^3 -3 n^2 + 2n$. Let us connect this point of view with the symmetry perspective above. We reintroduce the restriction \( i>j \) and \( k>l \) for \( c_i^{\dag}c_j^{\dag}c_k c_l \). There are four separate cases of coincidence of two indices with the other two distinct from each other: a) \( k=i \), and \( j\neq l \); b) \( l=i \), which implies \( k>i>j \); c) \( k=j \), leading to \( i>j>l \); and d) \( j=l \), with \( i\neq k \).  There are \( \binom{n}{3} \) cases of types b) and c), and \( 2\times \binom{n}{3}\) each for cases a) and d). All these cases can be unified in terms of the form \( c_i^{\dag}c_j^{\dag}c_j c_l \) (type (ii)) by relaxing the ordering condition. There are altogether, as before, \( 6\times \binom{n}{3}=n^3 -3 n^2 + 2n \) terms of this kind.

There is only one type of coincidence of two pairs, given by \( k=i \) and \( j=l \), with \( i>j \) (type (iii)), and there are \( \binom{n}{2} \) of these terms.

Altogether we see that there are \( \binom{n}{2}^2-6\binom{n}{3}- \binom{n}{2}=6\binom{n}{4} \) independent terms of type (i). This does not yet mean that their coefficients are independent stochastic variables. We have not imposed as yet hermiticity of the Hamiltonian.

There is a further simplification for interactions of type (i), in which the terms are grouped as
\begin{eqnarray}\label{eq:jletter}
J^1_{ij;k\ell} &=& \tilde{J}_{ij;k\ell} + \tilde{J}_{ik;j\ell} + \tilde{J}_{i\ell;jk} \ ,\nonumber \\
J^2_{ij;k\ell} &=& \tilde{J}_{ij;k\ell} - \tilde{J}_{ik;j\ell} + \tilde{J}_{i\ell;jk} \ ,\nonumber \\
J^3_{ij;k\ell} &=& \tilde{J}_{ij;k\ell} + \tilde{J}_{ik;j\ell} - \tilde{J}_{i\ell;jk} \ ,\nonumber \\
J^4_{ij;k\ell} &=& -\tilde{J}_{ij;k\ell} + \tilde{J}_{ik;j\ell} + \tilde{J}_{i\ell;jk} \ ,
\end{eqnarray}
which we notice  are related as $J^1_{ij;k\ell} = J^2_{ij;k\ell} + J^3_{ij;k\ell} + J^4_{ij;k\ell}$. We consider a larger number of dependent parameters in order to minimise spin interactions, as shown in Eq.~(\ref{G1}). The real coefficients appearing in the spin Hamiltonian are $\Re(J^{a}_{ij;k\ell})$ and $\Im(J^{a}_{ij;k\ell})$, with $a=1,2,3,4$.

As we must emphasise, the fermionic symmetries of these models  entail that not all coefficients are independent. In fact,  a choice of indices \( i>j>k>l \) determines all the possible six orderings of those distinct  numbers that maintain the property that both the first and the second pair are ordered, i.e. for type (i). This will give rise to the combinations above, in Eq.~(\ref{eq:jletter}). The complex coefficients $J_{ij;k\ell}$ in the fermionic  model satisfy a Gaussian random distribution with zero mean. If we consider that the real and imaginary parts are independently Gaussian distributed and have zero mean, then $\Re(J^{a}_{ij;k\ell})$ and $\Im(J^{a}_{ij;k\ell})$ with $a=2,3,4$, which are a linear combination of Gaussian stochastic variables are themselves Gaussian distributed. Assuming that in the initial definition of the complex coefficients th real and imaginary parts are identically distributed, so are $\Re(J^{a}_{ij;k\ell})$ and $\Im(J^{a}_{ij;k\ell})$, more concretely \( \sim N(0,6 \sqrt{2} J) \). Furthermore, they are independent when we set  \( i>j>k>l \), and identically distributed. We note that $J^{1}_{ij;k\ell}$ is obtained from the sum of $J^{2}_{ij;k\ell}$, $J^{3}_{ij;k\ell}$ and $J^{4}_{ij;k\ell}$. Summarising, for the case of distinct indices we have six independent identically distributed (i.i.d.) real Gaussian random variables for the \( \binom{n}{4} \) alternatives \( i>j>k>l \), starting from complex coefficients. If we were to restrict ourselves to real coefficients, there would be just three i.i.d. real Gaussian random variables for those alternatives, \( \sim N(0,12 J) \) .

\section{Construction of spin interaction terms for the quantum algorithm}

\subsection{Details on Majorana models}

As stated in the main text, the Jordan--Wigner construction for Majorana fermions is given by
\begin{equation}
\chi_l\to \left(\prod_{j=1}^{\tilde{l}-1} \sigma_j^z\right) \sigma_{\tilde{l}}^{ \alpha_l}\,,
\end{equation}
where \( \tilde{l}=\lfloor (l+1)/2\rfloor \) and \( \alpha_l \) is \( x \) for even \( l \) and \( y \) for odd \( l \).

The Majorana interaction terms of type (i), \( \chi_i \chi_j \chi_k \chi_l \) with \( i>j>k>l \), correspond to spin interaction terms of the form $\left(\prod_{m = \tilde{l}}^{\tilde{k}-1} \sigma^z_m \right) \left(\prod_{m = \tilde{j}}^{\tilde{i}-1} \sigma^z_m \right) \sigma^{\alpha_i}_{\tilde{i}} \sigma^{\alpha_j}_{\tilde{j}} \sigma^{\alpha_k}_{\tilde{k}} \sigma^{\alpha_l}_{\tilde{l}}$. Note that each fermionic interaction term only is translated into a single kind of spin interaction, with a definite combination of $\alpha$ spin indices. This general expression can be reduced by the following observation: a pair of ordered indices \(s \) and \( t \), with \( s>t \), give identical reduced indices \( \tilde{s} =\tilde{t}\) only if both a) \( s=t+1 \) and b) \( t \) is odd (\( s \) is even) are fulfilled. Therefore, there are four possibilities of a simplification in the spin presentation: A) \( \tilde{i}=\tilde{j}>\tilde{k}\geq\tilde{l} \); B) \( \tilde{i}\geq\tilde{j}>\tilde{k}=\tilde{l} \); C) \( \tilde{i}>\tilde{j}=\tilde{k}>\tilde{l} \). The possibility does exist that \( \tilde{i}=\tilde{j}>\tilde{k}=\tilde{l} \), which has been included in both A and B. We present these cases and the number of combinations corresponding to each in Table~\ref{Majorana fermionic model i} in terms of Majorana terms.

\begin{table}[h]
\centering
\caption{Majorana quartic fermionic interaction terms when $i > j > k > l$.}
\label{Majorana fermionic model i}
\renewcommand{\arraystretch}{1.5}
\begin{tabular}{|c|c|c|}
\hline
Set & Fermionic interaction & Number of terms \\  \hline 
$U$ & $\chi_i \chi_j \chi_k \chi_l$ & $|U| = \binom{2n}{4} = \frac{2}{3} n^4 -2 n^{3} + \frac{11}{6} n^{2} -\frac{1}{2} n$ \\  \hline 
$A$ & $\chi_i \chi_{i-1} \chi_k \chi_l$ $\forall i$ even & $|A| = \sum_{m=2}^n \sum_{k=2}^{2m-2} \sum_{l=1}^{k-1} 1 = \frac{2}{3} n^3 - \frac{3}{2} n^2 + \frac{5}{6} n$  \\  \hline 
$B$ & $\chi_i \chi_j \chi_k \chi_{k-1}$ $\forall k$ even & $|B| = \sum_{i=4}^{2n} \sum_{j=3}^{i-1} \left\lfloor \frac{j-1}{2}\right\rfloor = \frac{2}{3} n^3-\frac{3}{2} n^2+\frac{5}{6} n$ \\  \hline 
$A\cap B$ & $\chi_i \chi_{i-1} \chi_k \chi_{k-1}$ $\forall i,k$ even & $|A\cap B| =\binom{n}{2}= \frac{1}{2}n^2 - \frac{1}{2}n$  \\  \hline 
$C$ & $\chi_i \chi_j \chi_{j-1} \chi_l$ $\forall j$ even & $|C| = \sum_{i=5}^{2n} \sum_{m=2}^{\tilde{i}-1} \sum_{l=1}^{2m-2} 1 = \frac{2}{3} n^3 -2 n^2+\frac{4}{3} n $ \\  \hline
\end{tabular}
\end{table}

Now we state the simplifications in spin interaction terms as shown in Table~ \ref{Majorana spin model}. First, the cases \( A\setminus\left\{A\cap B\right\}\) and \( B\setminus\left\{A\cap B\right\} \) present a similar structure. As an example, the spin interaction term for \( A\setminus\left\{A\cap B\right\}\) becomes \( i \left(\prod_{m = \tilde{l}}^{\tilde{k}-1} \sigma^z_m \right) \sigma^{z}_{\tilde{i}} \sigma^{\alpha_k}_{\tilde{k}} \sigma^{\alpha_l}_{\tilde{l}} \). Notice the presence of the \( i \) factor, which is in fact required for hermiticity, because \( (\sigma_{\tilde{l}}^z\sigma_{\tilde{l}}^{\alpha_l})^{\dag}=- \sigma_{\tilde{l}}^z\sigma_{\tilde{l}}^{\alpha_l} \).

The intersection set \( A\cap B \) has  representative \( - \sigma^z_{\tilde{i}}\sigma^z_{\tilde{k}} \).

Finally, the central coincidence of the indices in terms belonging to set $C$ produces spin interaction terms of the form $i\left(\prod_{m = \tilde{l}}^{\tilde{k}-1} \sigma^z_m \right) \left(\prod_{m = \tilde{j}}^{\tilde{i}-1} \sigma^z_m \right) \sigma^{\alpha_i}_{\tilde{i}} \sigma^{z}_{\tilde{j}}  \sigma^{\alpha_l}_{\tilde{l}}$.

\begin{table}[h]
\centering
\caption{Spin interactions for quartic Majorana fermionic terms.}
\label{Majorana spin model}
\renewcommand{\arraystretch}{1.5}
\begin{tabular}{|c|c|c|}
\hline
Set & Spin interaction & Number of terms \\ \hline
$U \setminus \{A\cup B\cup C\}$ & $\left(\prod_{m = \tilde{l}}^{\tilde{k}-1} \sigma^z_m \right) \left(\prod_{m = \tilde{j}}^{\tilde{i}-1} \sigma^z_m \right) \sigma^{\alpha_i}_{\tilde{i}} \sigma^{\alpha_j}_{\tilde{j}} \sigma^{\alpha_k}_{\tilde{k}} \sigma^{\alpha_l}_{\tilde{l}}$ & $\frac{2}{3} n^4 -4 n^3+\frac{22}{3} n^2-4 n$  \\ \hline
$A \setminus \{A\cap B\}$ & $i \left(\prod_{m = \tilde{l}}^{\tilde{k}-1} \sigma^z_m \right) \sigma^{z}_{\tilde{i}} \sigma^{\alpha_k}_{\tilde{k}} \sigma^{\alpha_l}_{\tilde{l}}$ & $\frac{2}{3} n^3-2 n^2+\frac{4}{3} n$  \\ \hline
$B \setminus \{A\cap B\}$ & $i\left(\prod_{m = \tilde{j}}^{\tilde{i}-1} \sigma^z_m \right) \sigma^{\alpha_i}_{\tilde{i}} \sigma^{\alpha_j}_{\tilde{j}} \sigma^{z}_{\tilde{k}}$ & $\frac{2}{3} n^3-2 n^2+\frac{4}{3} n$ \\ \hline
$A\cap B$ & -$\sigma^{z}_{\tilde{i}} \sigma^{z}_{\tilde{k}}$ & $\frac{n^2}{2}-\frac{n}{2}$  \\ \hline
$C$ & $i\left(\prod_{m = \tilde{l}}^{\tilde{k}-1} \sigma^z_m \right) \left(\prod_{m = \tilde{j}}^{\tilde{i}-1} \sigma^z_m \right) \sigma^{\alpha_i}_{\tilde{i}} \sigma^{z}_{\tilde{j}}  \sigma^{\alpha_l}_{\tilde{l}}$ & $\frac{2}{3} n^3 -2 n^2+\frac{4}{3} n$ \\ \hline
\end{tabular}
\end{table}

Coming now to quadratic terms \( \chi_i \chi_j \), with \( i>j \), notice again that they are antihermitian as such. They are generically codified as 
\begin{equation}
i \chi_i \chi_j\to i\left(\prod_{l=\tilde{j}}^{\tilde{i}-1} \sigma_l^z\right) \sigma_{\tilde{i}}^{\alpha_i}\sigma_{\tilde{j}}^{\alpha_j}\,,
\end{equation}
and there are \( \binom{2n}{2} \) of them. The special case is here when \( i \) is even and \( j=i-1 \). There are \( n \) possibilities, all of which are represented  by \( - \sigma_{\tilde{i}}^z \).

\subsection{Details on spinless complex fermion models}
Let us now consider the model with complex spinless fermions. The interaction terms can be mapped as above to spin interactions via the Jordan--Wigner transformation. Thus, the interaction terms of type (i) of this model are expressed as
\begin{equation}
c_i^\dagger c_j^\dagger c_k^{\vphantom \dagger} c_\ell^{\vphantom \dagger} = \beta \left(\prod_{\xi= \zeta_1 +1}^{\zeta_2 -1} \sigma^z_\xi \right) \left(\prod_{\xi= \zeta_3 +1}^{\zeta_4 -1} \sigma^z_\xi \right) \sigma^+_i \sigma^+_j \sigma^-_k \sigma^-_\ell \ ,
\end{equation}
where $\{\zeta_1,\zeta_2,\zeta_3,\zeta_4\}$ is the permutation of $\{i,j,k,\ell\}$ such that $\zeta_1< \zeta_2< \zeta_3<\zeta_4$, and $\beta= \text{sign}(i-j) \text{sign}(\ell-k)$. We remark that this general expression for distinct indices holds for any ordering of them. Now, in the quantum simulation, we focus on the terms such that $i>j$ and $k>l$, for which $\beta = -1$ in our case for type (i).

Let us begin with the general case of complex coefficients $J_{ij;k\ell}$. The complex fermionic interaction terms with all indices distinct from each other, can be rearranged such that the interaction coefficients for spins are real. By means of the coefficient relation $J_{k\ell;ij} = J_{ij;k\ell}^\ast$, and the identity $\sigma^{\pm}=(\sigma^{x} \pm i \sigma^{y})/2$, the corresponding spin interaction terms read
\begin{eqnarray}
J_{ij;k\ell} c_i^\dagger c_j^\dagger c_k^{\vphantom \dagger} c_\ell^{\vphantom \dagger} &+& J_{k\ell;ij} c_k^\dagger c_\ell^\dagger c_i^{\vphantom \dagger} c_j^{\vphantom \dagger} = \frac{\beta}{8} \left(\prod_{\xi= \zeta_1 +1}^{\zeta_2 -1} \sigma^z_\xi \right) \left(\prod_{\xi= \zeta_3 +1}^{\zeta_4 -1} \sigma^z_\xi \right) \nonumber \\
& & \cdot \Bigg[ \Re(J_{ij;k\ell}) ( \sigma^x_i \sigma^x_j \sigma^x_k \sigma^x_\ell + \sigma^y_i \sigma^y_j \sigma^y_k \sigma^y_\ell + \sigma^y_i \sigma^x_j \sigma^y_k \sigma^x_\ell + \sigma^x_i \sigma^y_j \sigma^x_k \sigma^y_\ell \nonumber \\
& & + \sigma^y_i \sigma^x_j \sigma^x_k \sigma^y_\ell + \sigma^x_i \sigma^y_j \sigma^y_k \sigma^x_\ell - \sigma^x_i \sigma^x_j \sigma^y_k \sigma^y_\ell - \sigma^y_i \sigma^y_j \sigma^x_k \sigma^x_\ell) \nonumber \\ [10pt]
& & + \Im(J_{ij;k\ell}) ( \sigma^y_i \sigma^x_j \sigma^x_k \sigma^x_\ell - \sigma^x_i \sigma^x_j \sigma^y_k \sigma^x_\ell +\sigma^x_i \sigma^y_j \sigma^x_k \sigma^x_\ell - \sigma^x_i \sigma^x_j \sigma^x_k \sigma^y_\ell \nonumber \\ 
& & + \sigma^y_i \sigma^y_j \sigma^x_k \sigma^y_\ell - \sigma^x_i \sigma^y_j \sigma^y_k \sigma^y_\ell + \sigma^y_i \sigma^y_j \sigma^y_k \sigma^x_\ell - \sigma^y_i \sigma^x_j \sigma^y_k \sigma^y_\ell) \Bigg] \ .
\end{eqnarray}
Notice that this relation is valid for all orderings of indices, and that for $\zeta_2-\zeta_1=1$ and $\zeta_4-\zeta_3=1$ the products are substituted by the identity operator. We will take into account the model rewritten in terms of $\tilde{J}_{\alpha_{2}\alpha_1;\beta_2\beta_1}$ for interaction terms with distinct indices. In this case, by definition, those coefficients are different from zero only for $\alpha_2>\alpha_1$ and $\beta_2>\beta_1$, for which $\beta = -1$. That is, it accounts for those interactions classified as type (i) in the text. We now want to identify a minimal set of spin interaction terms with real coefficients. We thus group type (i) complex fermionic terms by making use of the previous relation of conjugated interactions. Now, we choose a set of ordered indices \( i>j>k>l \). This determines six orderings  \( \alpha_1\alpha_2;\beta_1 \beta_2 \) which respect $\alpha_2>\alpha_1$ and $\beta_2>\beta_1$, namely, \( \left\{(ij;kl),(kl;ij),(ik;jl),(jl;ik),(il;jk),(jk;il)\right\} \). Then, keeping this choice \( i>j>k>l \), we group 
\begin{eqnarray}
\label{G1}
&&\tilde{J}_{ij;k\ell} c_i^\dagger c_j^\dagger c_k^{\vphantom \dagger} c_\ell^{\vphantom \dagger} + \tilde{J}_{k\ell;ij} c_k^\dagger c_\ell^\dagger c_i^{\vphantom \dagger} c_j^{\vphantom \dagger} + \tilde{J}_{ik;j\ell} c_i^\dagger c_k^\dagger c_j^{\vphantom \dagger} c_\ell^{\vphantom \dagger} + \tilde{J}_{j\ell;ik} c_j^\dagger c_\ell^\dagger c_i^{\vphantom \dagger} c_k^{\vphantom \dagger} + \tilde{J}_{i\ell;jk} c_i^\dagger c_\ell^\dagger c_j^{\vphantom \dagger} c_k^{\vphantom \dagger} + \tilde{J}_{jk;i\ell} c_j^\dagger c_k^\dagger c_i^{\vphantom \dagger} c_\ell^{\vphantom \dagger} = \nonumber \\ [10pt]
&& \quad \frac{\beta}{8} \left(\prod_{\xi= \zeta_1 +1}^{\zeta_2 -1} \sigma^z_\xi \right) \left(\prod_{\xi= \zeta_3 +1}^{\zeta_4 -1} \sigma^z_\xi \right) \cdot \Bigg[ \Re(\tilde{J}_{ij;k\ell} + \tilde{J}_{ik;j\ell} + \tilde{J}_{i\ell;jk}) ( \sigma^x_i \sigma^x_j \sigma^x_k \sigma^x_\ell + \sigma^y_i \sigma^y_j \sigma^y_k \sigma^y_\ell) \nonumber \\ 
&& \quad + \Re(\tilde{J}_{ij;k\ell} - \tilde{J}_{ik;j\ell} + \tilde{J}_{i\ell;jk}) (\sigma^y_i \sigma^x_j \sigma^y_k \sigma^x_\ell + \sigma^x_i \sigma^y_j \sigma^x_k \sigma^y_\ell) + \Re(\tilde{J}_{ij;k\ell} + \tilde{J}_{ik;j\ell} - \tilde{J}_{i\ell;jk}) (\sigma^y_i \sigma^x_j \sigma^x_k \sigma^y_\ell + \sigma^x_i \sigma^y_j \sigma^y_k \sigma^x_\ell) \nonumber \\ [10pt]
&& \quad + \Re(-\tilde{J}_{ij;k\ell} + \tilde{J}_{ik;j\ell} + \tilde{J}_{i\ell;jk}) (\sigma^x_i \sigma^x_j \sigma^y_k \sigma^y_\ell + \sigma^y_i \sigma^y_j \sigma^x_k \sigma^x_\ell) + \Im(\tilde{J}_{ij;k\ell} + \tilde{J}_{ik;j\ell} + \tilde{J}_{i\ell;jk}) (\sigma^y_i \sigma^x_j \sigma^x_k \sigma^x_\ell - \sigma^x_i \sigma^y_j \sigma^y_k \sigma^y_\ell)\nonumber \\ [10pt]
&& \quad + \Im(\tilde{J}_{ij;k\ell} - \tilde{J}_{ik;j\ell} + \tilde{J}_{i\ell;jk}) (\sigma^y_i \sigma^y_j \sigma^x_k \sigma^y_\ell - \sigma^x_i \sigma^x_j \sigma^y_k \sigma^x_\ell) + \Im(\tilde{J}_{ij;k\ell} + \tilde{J}_{ik;j\ell} - \tilde{J}_{i\ell;jk})(\sigma^y_i \sigma^y_j \sigma^y_k \sigma^x_\ell - \sigma^x_i \sigma^x_j \sigma^x_k \sigma^y_\ell)\nonumber \\ 
&& \quad + \Im(-\tilde{J}_{ij;k\ell} + \tilde{J}_{ik;j\ell} + \tilde{J}_{i\ell;jk}) (\sigma^y_i \sigma^x_j \sigma^y_k \sigma^y_\ell - \sigma^x_i \sigma^y_j \sigma^x_k \sigma^x_\ell)\Bigg] = \nonumber \\
&& \quad \frac{\beta}{8} \left(\prod_{\xi= \zeta_1 +1}^{\zeta_2 -1} \sigma^z_\xi \right) \left(\prod_{\xi= \zeta_3 +1}^{\zeta_4 -1} \sigma^z_\xi \right) \cdot \Bigg[ \Re(J^1_{ij;k\ell}) ( \sigma^x_i \sigma^x_j \sigma^x_k \sigma^x_\ell + \sigma^y_i \sigma^y_j \sigma^y_k \sigma^y_\ell) + \Re(J^2_{ij;k\ell}) (\sigma^y_i \sigma^x_j \sigma^y_k \sigma^x_\ell + \sigma^x_i \sigma^y_j \sigma^x_k \sigma^y_\ell) \nonumber \\ 
&& \quad + \Re(J^3_{ij;k\ell}) (\sigma^y_i \sigma^x_j \sigma^x_k \sigma^y_\ell + \sigma^x_i \sigma^y_j \sigma^y_k \sigma^x_\ell) + \Re(J^4_{ij;k\ell}) (\sigma^x_i \sigma^x_j \sigma^y_k \sigma^y_\ell + \sigma^y_i \sigma^y_j \sigma^x_k \sigma^x_\ell) + \Im(J^1_{ij;k\ell}) (\sigma^y_i \sigma^x_j \sigma^x_k \sigma^x_\ell - \sigma^x_i \sigma^y_j \sigma^y_k \sigma^y_\ell) \nonumber \\ 
&& \quad + \Im(J^2_{ij;k\ell}) (\sigma^y_i \sigma^y_j \sigma^x_k \sigma^y_\ell - \sigma^x_i \sigma^x_j \sigma^y_k \sigma^x_\ell) + \Im(J^3_{ij;k\ell})(\sigma^y_i \sigma^y_j \sigma^y_k \sigma^x_\ell - \sigma^x_i \sigma^x_j \sigma^x_k \sigma^y_\ell) + \Im(J^4_{ij;k\ell}) (\sigma^y_i \sigma^x_j \sigma^y_k \sigma^y_\ell - \sigma^x_i \sigma^y_j \sigma^x_k \sigma^x_\ell)\Bigg]. \nonumber \\
\end{eqnarray}

In principle, we have mapped fermionic interaction terms with 6 independent real coefficients in terms of spin interaction terms with 8 real coefficients. We have preferred to minimise the number of spin interactions, even if it implies using the dependent coefficients $J^{a}_{ij;k\ell}$ with $a=1,2,3,4$ defined previously. 

For interaction terms of type (ii), we do not impose the constraints $i>j$ and $k>l$, as can be seen in the classification. The complex interaction terms can be also reordered such that the interaction coefficients for spins are real,
\begin{equation}
J_{ij;jk} c_i^\dagger c_j^\dagger c_j^{\vphantom \dagger} c_k^{\vphantom \dagger} + J_{jk;ij} c_j^\dagger c_k^\dagger c_i^{\vphantom \dagger} c_j^{\vphantom \dagger} = -\frac{1}{4} \left(\prod_{\xi=\zeta_1+1}^{\zeta_2-1} \sigma^z_\xi \right)(\sigma^z_j +1) \Bigg[ \Re(J_{ij;jk}) ( \sigma^x_i \sigma^x_k + \sigma^y_i \sigma^y_k ) + \Im(J_{ij;jk}) (\sigma^x_i \sigma^y_k - \sigma^y_i \sigma^x_k) \Bigg] \ .
\end{equation}

We note that interactions (iii) and (iv) can be grouped as follows
\begin{equation}
4 J_{ij;ji} n_i n_j + \mu n_i + \mu n_j =  \left(J_{ij;ji}+\frac{\mu}{2}\right)  \left( 2 + \sigma^{z}_i + \sigma^{z}_j \right) + J_{ij;ji} \sigma^{z}_i \sigma^{z}_j 
\end{equation}
so that we only take into account the interaction terms of type (iii) in the counting, where $J_{ij;ji}$ must be real because of the symmetries.

We show the counting of complex fermionic interaction terms in Table~\ref{Complex fermionic int terms}, and the corresponding spin interactions in Table~\ref{Complex spin model}.

\begin{table}[h]
\centering
\caption{Complex fermionic interaction terms.}
\label{Complex fermionic int terms}
\renewcommand{\arraystretch}{1.5}
\begin{tabular}{|c|c|}
\hline
Fermionic interaction & Number of terms \\  \hline 
$J_{ij;k\ell} c_i^\dagger c_j^\dagger c_k^{\vphantom \dagger} c_\ell^{\vphantom \dagger} + J_{k\ell;ij} c_k^\dagger c_\ell^\dagger c_i^{\vphantom \dagger} c_j^{\vphantom \dagger}$ & $\frac{1}{2} \binom {n} {2} \binom {n-2} {2}= \frac{1}{8}n^4 -\frac{3}{4} n^3 +\frac{11}{8}n^2 -\frac{3}{4}n$ \\  \hline 
$J_{ij;jk} c_i^\dagger n_j c_k^{\vphantom \dagger} + J_{jk;ij} c_k^\dagger n_j c_i^{\vphantom \dagger}$ & $\frac{1}{2}n^3 -\frac{3}{2}n^2 + n$  \\  \hline 
$n_i n_j$ & $\frac{1}{2}n^2 -\frac{1}{2}n$ \\  \hline 
\end{tabular}
\end{table}

\begin{table}[h]
\centering
\caption{Complex fermionic interaction terms and their corresponding spin interaction translation.}
\label{Complex spin model}
\renewcommand{\arraystretch}{1.5}
\begin{tabular}{|c|c|}
\hline
Spin interaction & Number of spin terms \\ \hline
$\left(\prod_{\xi= \zeta_1 +1}^{\zeta_2 -1} \sigma^z_\xi \right) \left(\prod_{\xi= \zeta_3 +1}^{\zeta_4 -1} \sigma^z_\xi \right) \sigma^{\alpha_i}_i \sigma^{\alpha_j}_j \sigma^{\alpha_k}_k \sigma^{\alpha_l}_l$ & $\frac{16}{6} \binom {n} {2} \binom {n-2} {2} = \frac{2}{3}n^4 - 4 n^3 + \frac{22}{3} n^2 - 4n$ \\ \hline
$\left(\prod_{\xi=\zeta_1+1}^{\zeta_2-1} \sigma^z_\xi \right)(\sigma^z_j +1)\sigma^{\alpha_i}_i \sigma^{\alpha_k}_k$ & $\frac{4}{2} \frac{n!}{(n-3)!} = 2 n^3 -6 n^2 + 4n$ \\ \hline
$\sigma^{z}_i + \sigma^{z}_j + \sigma^{z}_i \sigma^{z}_j$ & $\binom {n} {2} = \frac{1}{2} n^2 -\frac{1}{2} n$ \\ \hline
\end{tabular}
\end{table}

\section{Gate count comparison per Trotter step}
We compare the number of resulting spin interactions for all models. Here, we denote as first Majorana model that with only quartic interaction terms; as second Majorana model, that which also includes quadratic terms; as first complex model, that with complex coupling constants for complex fermionic interactions; and as second complex model, the restriction of the previous one to only real coupling constants. 

The number of spin interactions after the Jordan--Wigner mapping is directly related with the number of gates needed for the quantum simulation. For trapped ions, one needs $O(1)$ gates per interaction term, whereas it scales with $O(N)$ per interaction for the decomposition of the algorithm in single- and two-qubit gates in superconducting circuits.

In principle, we have found that the second complex model is more suitable to be simulated with our method due to the total amount of gates required. We have not taken into account the length of the Jordan--Wigner strings, $\prod_\xi \sigma^{z}_\xi$, but only treated them as an element of multiqubit gates.

We analyse deeper these contributions, by decomposing all kind of multiqubit gates into two-qubit and single-qubit gates for all the models that we have considered, and show the counting in Table~\ref{Complex spin model ii}.

\begin{table}[ht]
\centering
\caption{Independent spin interactions for all models.}
\label{Complex spin model ii}
\renewcommand{\arraystretch}{1.5}
\begin{tabular}{|c|c|c|c|c|}
\hline
Gates & First Majorana model  & Second Majorana model & First complex model & Second complex model \\ \hline
$\left(\prod_\xi \sigma^z_\xi \right) \left(\prod_\xi \sigma^z_\xi \right) \sigma^{\alpha_i}_i \sigma^{\alpha_j}_j \sigma^{\alpha_k}_k \sigma^{\alpha_l}_l$ &  $\frac{2}{3} n^4 -4 n^3+\frac{22}{3} n^2-4 n$ & $\frac{2}{3} n^4 -4 n^3+\frac{22}{3} n^2-4 n$ & $\frac{2}{3}n^4 - 4 n^3 + \frac{22}{3} n^2 - 4n$ & $\frac{1}{3}n^4 - 2 n^3 + \frac{11}{3} n^2 - 2n$ \\ \hline
$\left(\prod_\xi \sigma^z_\xi \right) \left(\prod_\xi \sigma^z_\xi \right) \sigma^{\alpha_i}_i \sigma^{\alpha_j}_j \sigma^z_k$ & $\frac{2}{3} n^3 -2 n^2+\frac{4}{3} n$ & $\frac{2}{3} n^3 -2 n^2+\frac{4}{3} n$ & None & None \\ \hline
$\left(\prod_\xi \sigma^z_\xi \right) \sigma^{\alpha_i}_i \sigma^{\alpha_j}_j \sigma^{z}_k$ & $\frac{4}{3} n^3-4 n^2+\frac{8}{3} n$ & $\frac{4}{3} n^3-4 n^2+\frac{8}{3} n$ & $2n^3 - 6n^2 +4n$ & $n^3 - 3n^2 +2n$ \\ \hline
$\left(\prod_\xi \sigma^z_\xi \right)\sigma^{\alpha_i}_i \sigma^{\alpha_k}_k$ & None & $2n^2-2n$ & $2n^2-2n$ & $n^2-n$ \\ \hline
$\sigma^{z}_i \sigma^{z}_j$ & $\frac{1}{2}n^2 -\frac{1}{2}n$ & $\frac{1}{2}n^2 -\frac{1}{2}n$ & $\frac{1}{2} n^2 -\frac{1}{2} n$ & $\frac{1}{2} n^2 -\frac{1}{2} n$ \\ \hline
$\sigma^{z}_i$ & None & $n$ & $n$ & $n$ \\ \hline
\end{tabular}
\end{table}

\end{widetext}

\end{document}